\documentclass[twosided,12pt]{article}

\usepackage{graphicx}

\newcommand{\be}{\begin{equation}}
\newcommand{\ee}{\end{equation}}
\newcommand{\bea}{\begin{eqnarray}}
\newcommand{\eea}{\end{eqnarray}}

\begin{document}

\title{ \vspace{1cm} The ``Perfect'' Fluid Quenches Jets Almost Perfectly}
\author{Berndt M\"uller\\
\\
Department of Physics, Duke University, Durham, NC 27708-0305, USA}
\maketitle
\begin{abstract} 
The QCD matter produced in nuclear collisions at the Relativistic Heavy Ion Collider (RHIC) has been found to have a very low shear viscosity, which is close to the lower bound allowed by unitarity. The matter has also been found to strongly suppress the emission of energetic hadrons. This phenomenon, called ``jet quenching'' is interpreted to be the result of a large energy loss by the precursor parton on its path through the dense matter, primarily due to gluon radiation. I discuss how the two phenomena are related. The RHIC data suggest, in some scenarios of jet quenching, that the quark-gluon plasma created in nuclear collisions is characterized by strong coupling, but still admits a quasi-particle description.
\end{abstract}


\section{Introduction}

The Relativistic Heavy Ion Collider (RHIC) was constructed to explore the physics of the quark-gluon plasma, a novel state of nuclear matter in which quarks and gluons are not locally confined to color singlet bound states \cite{Harris:1996zx}. The experiments performed at the RHIC facility since the year 2000 have confirmed \cite{Arsene:2004fa,Back:2004je,Adams:2005dq,Adcox:2004mh} that thermalized matter with an initial energy density exceeding 5 GeV/fm$^3$ (more than 30 times normal nuclear matter density) is created in collisions of two heavy nuclei at a center-of-mass energy per nucleon pair $\sqrt{s_{\rm NN}} = 200$ GeV. The final state is well described in its chemical composition as a hadron gas with temperature $T_\textrm{ch} \approx 160$ MeV \cite{Kaneta:2004zr,Andronic:2006ky}, expanding with a collective velocity exceeding half the speed of light. Two primary probes of the properties of this matter before its dissociation into individual hadrons have emerged from the RHIC experiments: 
\begin{enumerate}
\item The axial anisotropy of the collective flow pattern (called ``elliptic flow'') of the final-state hadrons with respect to the beam axis in off-central collisions;
\item The suppression of the emission of hadrons with a high transverse momentum $p_T$, compared with the yield expected from a superposition of independent nucleon-nucleon interactions.
\end{enumerate}
Theoretical studies, discussed below, have related each of these phenomena to a transport coefficient of the produced matter: The elliptic flow has been shown to be highly sensitive to the value of the shear viscosity $\eta$; the energy loss responsible for the high-$p_T$ hadron suppression has been shown to be governed a parameter $\hat{q}$, which specifies the average square of the transverse momentum kicks per unit path length experienced by a parton traversing the matter. Remarkably, it turns out that these two parameters are related to each other, if the matter is a quasi-particle medium composed of quasi-particle excitations with the quantum numbers of quarks and gluons, i.~e.\ a {\em quark-gluon plasma} in the literal sense of the word \cite{Majumder:2007zh}. On the other hand, if the medium is so strongly coupled that  no quasi-particle description applies, or if the quasi-particles have different quantum numbers than quarks and gluons, the relation fails to hold. A reliable measurement of both parameters, $\eta$ and $\hat{q}$, would enable us to differentiate between such different scenarios.

\section{Shear Viscosity and Elliptic Flow}

When two heavy nuclei collide off-centrally, i.~e.\ with a nonzero impact parameter $b$, the overlap region is shaped roughly like an American football. In a plane transverse to the beam axis, the region is almond-shaped and can be characterized by a spatial eccentrity $\varepsilon$. When such a region is filled with equilibrated matter endowed with a large internal pressure, its geometrical shape ensures that the pressure gradient in the direction of the collision plane is larger than the gradient perpendicular to that plane. This pressure anisotropy leads to a more violent expansion along the collision plane and thus results in an axially anisotropic transverse (with respect to the beam axis) flow pattern, which is commonly characterized by the second Fourier moment $v_2$ of the angular distribution of hadrons for a given $p_T$. The elliptic flow parameter $v_2$ can be measured as a function of $p_T$ and also for individual hadron species.

\begin{figure}[tb]
\begin{center}
\begin{minipage}[b]{0.4\textwidth}
\includegraphics[width=\linewidth]{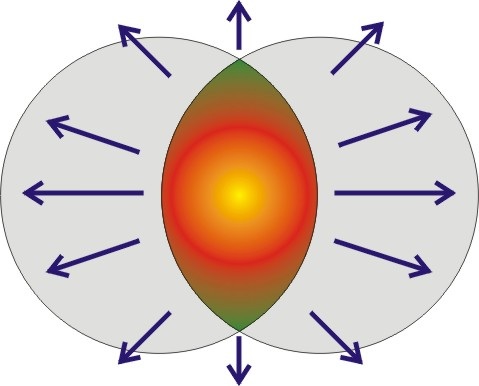}
\end{minipage}
\hskip 0.05\textwidth
\begin{minipage}[b]{0.45\textwidth}
\caption{In an off-central collision between two heavy nuclei, the overlap region is almond-shaped and can be characterized by a spatial eccentrity $\varepsilon$. The large pressure gradient in the horizontal direction (the collision plane) causes a faster expansion than in the vertical direction. The momentum anisotropy is characterised by the elliptic flow parameter $v_2$.
\label{fig1}}
\end{minipage}
\end{center}
\end{figure}

It is not surprising that the elliptic flow produced by a given geometric eccentricity is related by the ability of the matter to flow freely locally. In fluid dynamics, this ability is parametrized by the {\em shear viscosity} $\eta$. The shear viscosity describes the response of the medium to flow gradients. A low value of $\eta$ implies that the matter is insensitive to flow gradients. Relevant scales determining the importance of shear viscosity are: the {\em mean free path} $\lambda_f$, the {\em homogeneity length} $L$ of the flow pattern, and possibly the thermal wave length $\lambda_{\rm th}=T^{-1}$. The ratio between the mean free path and the homogeneity scale is called the {\em Knudsen number}: $K = \lambda_f/L$. When $K\ll 1$, the fluid can be described by viscous hydrodynamics; when $K \ge 1$ such a macroscopic description fails, and the fluid has to be described microscopically, e.~g.\ by the Boltzmann equation. 

\begin{figure}[tb]
\begin{center}
\begin{minipage}[b]{0.4\textwidth}
\includegraphics[width=\linewidth]{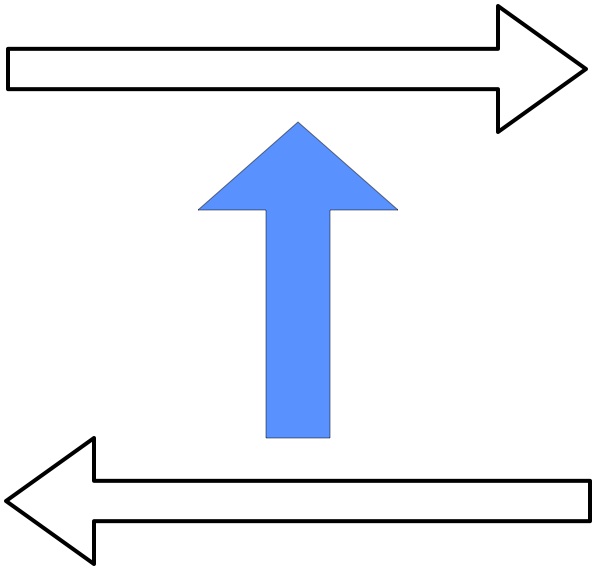}
\end{minipage}
\hskip 0.05\textwidth
\begin{minipage}[b]{0.45\textwidth}
\caption{Shear viscosity $\eta$ is caused by the exchange of momentum (vertical [blue] arrow) between regions with different flow velocity (horizontal arrows). A small shear viscosity implies a short mean free path for momentum transport in the fluid.
\label{fig2}}
\end{minipage}
\end{center}
\end{figure}

When there is a flow shear, i.~e.\ when the flow velocity changes in a direction transverse to the direction of the flow (see Fig.~\ref{fig2}), the shear viscosity is proportional to the distance over which longitudinal momentum can be readily transported transverse to the flow. Kinetic theory for a relativistic fluid yields the relation ($n$ is the particle density, $s$ the entropy density, $\bar{p}$ the average momentum of a fluid particle)
\begin{equation}
\eta \approx \frac{4}{15} n \bar{p} \lambda_f 
= \frac{4\, T}{5\, \sigma_\textrm{tr}}
\approx \frac{s}{5} T \lambda_f 
\approx \frac{s}{5}\,\frac{L}{\lambda_{\rm th}}\, K .
\end{equation}
In other words, a short mean free path or, equivalently, a large transport cross section $\sigma_\textrm{tr}$ imply a small shear viscosity. Somewhat counterintuitively, a low shear viscosity thus requires a strongly coupled medium! On the other hand, a strongly coupled medium can, and often does, have a large shear viscosity, because its degrees of freedom rearrange themselves under the strong interaction, and momentum can be transferred by mechanisms other than particle transport, e.~g.\ in solids by phonons, or in polymers by momentum transport along molecular chains. 

As a result, one finds in most fluids that the dimensionless ratio of shear viscosity to entropy density (a generalization of the concept of {\em kinematic viscosity})  first decreases rapidly with temperature, corresponding to an increase of the ratio between potential and kinetic energy, and then increases again below a critical temperature, often coinciding with a phase transition \cite{Chen:2007jq}, due to the emergence of new momentum transport mechanisms. An exception are scale invariant theories, such as the celebrated supersymmetric SU($N_c$) gauge theory at large $N_c$, for which the ratio $\eta/s$ continually decreases with increasing interaction strength and asymptotically reaches the value $1/4\pi$ \cite{Policastro:2001yc}, which is widely believed to be the lower bound allowed by quantum mechanics and unitarity \cite{Kovtun:2004de}. 

The question is sometimes raised whether hydrodynamic flow implies local equilibration of the medium. It is certainly true that a hydrodynamical description can be applied whenever a medium is locally thermalized and the Knudsen number is much less than unity. However, the reverse is not true! For example, collective motion in a plasma can be enforced by the action of magnetic fields, which effectively limit the ability of particle to move freely and transport momentum over large distances. This effect is encoded in the theory of {\em magneto-hydrodynamics}. Another role of magnetic fields in a plasma is their ability to reduce the shear viscosity by a large factor. Plasmas containing spontaneously generated, random magnetic fields are called ``turbulent'' plasmas. Similar mechanisms have been theoretically shown to occur in locally anisotropic quark-gluon plasmas \cite{Mrowczynski:1993qm,Rebhan:2005re,Arnold:2004ti}, and these may give rise to an anomalously low shear viscosity \cite{Asakawa:2006tc,Asakawa:2006jn}. Strong gluon fields are also thought to be present in the pre-thermal ``glasma'' phase and may be responsible for a small shear viscosity at early times.

As mentioned before, the shear viscosity of the quark-gluon plasma produced in nuclear collisions at RHIC can be constrained by a comparison of the measured elliptic flow parameter $v_2(p_T)$ with the results obtained by simulations of the transverse expansion in viscous hydrodynamics. The result of such comparisons \cite{Teaney:2003kp,Luzum:2008cw,Song:2007ux} is that the shear viscosity must be quite small, most likely in the range $0 < \eta/s < 0.3$. The extracted value depends somewhat on the initial conditions for the geometric anisotropy of the fireball used in the hydrodynamics calculations, with the Glauber model giving a value near the lower bound and parton saturation models giving a value near the upper bound. A slightly different way of extracting the value of $\eta/s$  is by comparing the hydrodynamics simulations with the dependence of the momentum averaged $v_2$ with the data for the average $v_2$ as a function of the charged particle multiplicity $dN/dy$ \cite{Song:2008si}. This comparison yields the range $0.15 \leq \eta/s \leq 0.3$.

\section{Radiative Energy Loss and Relation to Viscosity}

\begin{figure}[tb]
\begin{center}
\begin{minipage}[b]{0.4\textwidth}
\includegraphics[width=\linewidth]{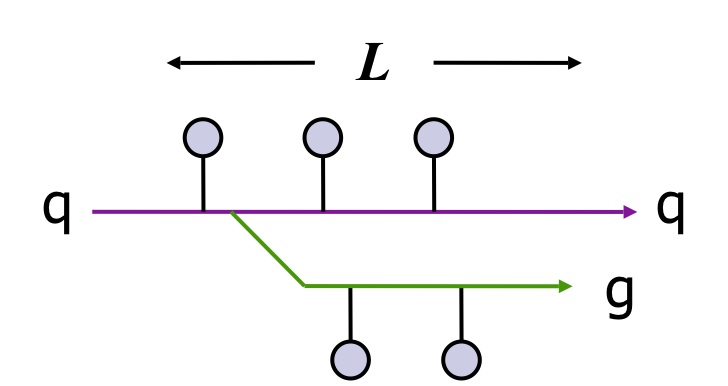}
\end{minipage}
\hskip 0.05\textwidth
\begin{minipage}[b]{0.45\textwidth}
\caption{An energetic parton traversing the quark-gluon plasma receives momentum kicks due to interactions with the medium. As a result, it radiates a fraction of its energy in form of gluons, which also interact with the medium. The transport coefficient $\hat{q}$ controlling the transverse momentum exchange determines the rate of radiative energy loss of the parton.
\label{fig3}}
\end{minipage}
\end{center}
\end{figure}

The dominant source of energy loss of an energetic parton traversing a quark-gluon plasma is believed to be gluon bremsstrahlung. The parton radiates away part of its energy after scattering off mass shell on one of the medium constituents (also partons if the quark-gluon plasma can be described as a quasi-particulate medium). In QCD, the energy loss can be expressed in terms of a parameter $\hat{q}$, which describes the rate at which the medium transfers transverse momentum to the fast parton. If the differential cross section for elastic scattering in the medium is given by $d\sigma/dk_T^2$, the energy loss parameter is given by \cite{Baier:1996kr}
\begin{equation}
\hat{q} = \rho \int k_T^2\, dk_T^2 \frac{d\sigma}{dk_T^2} ,
\end{equation}
and the energy loss per unit path length is $dE/dx = - \hat{q}L$, where $L$ denotes the total length of the path in the medium. The reason for the additional length dependence on the right-hand side of this equation is that also the radiated gluon re-scatters in the medium (see Fig.~\ref{fig3}). The quadratic path length dependence of the energy loss is characteristic of QCD.

\begin{figure}[tb]
\begin{center}
\begin{minipage}[b]{0.4\textwidth}
\includegraphics[width=\linewidth]{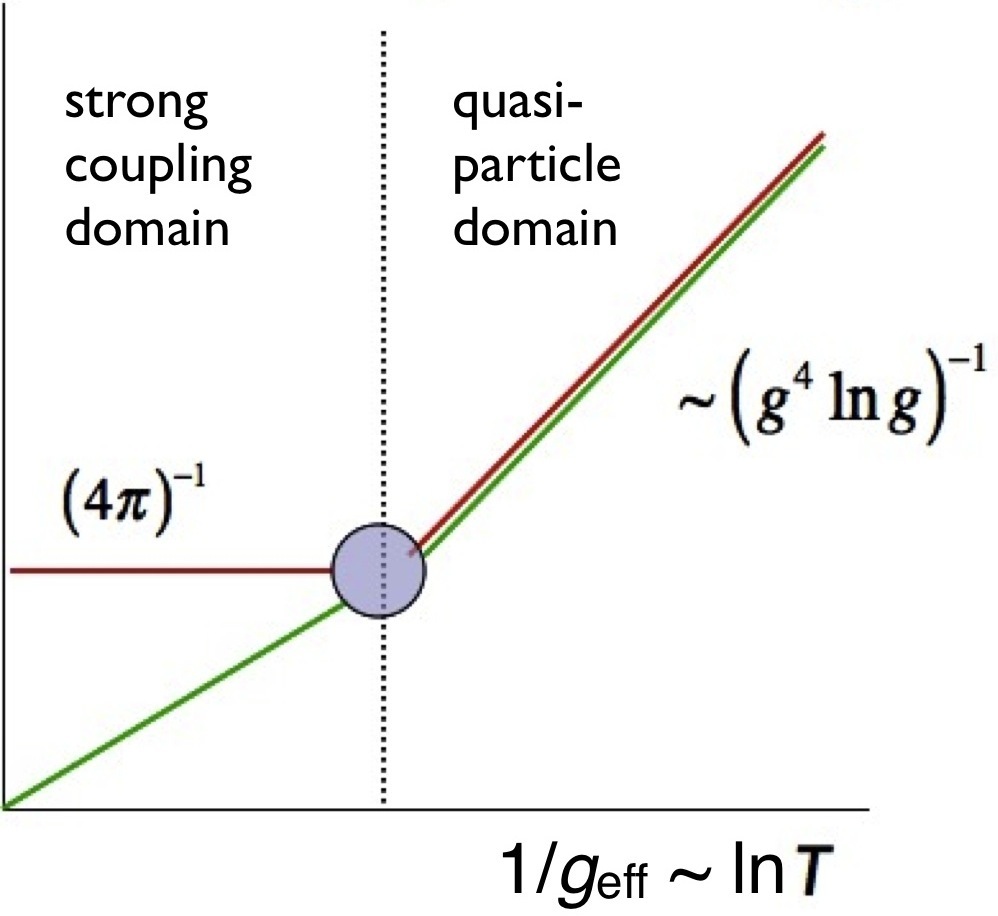}
\end{minipage}
\hskip 0.05\textwidth
\begin{minipage}[b]{0.45\textwidth}
\caption{$\eta/s$ (upper [red] line) and $1.25\,T^3/\hat{q}$ (lower [green] line) as function of the inverse coupling strength. The two quantities approximately agree in the weak coupling domain, when a quasi-particle picture applies, but not in the strong coupling domain, where $\eta/s > 1.25\,T^3/\hat{q}$.
\label{fig4}}
\vskip 0.5cm
\end{minipage}
\end{center}
\end{figure}

The fact that the radiative energy loss depends on the rate of momentum exchange between the fast parton and the medium allows us to relate the coefficient $\hat{q}$ to the shear viscosity, if the average thermal parton can be described as a quasi-particle with the same quantum numbers as a hard parton, i.e., a quark or gluon. We can then apply kinetic theory as discussed in the previous section and relate $\eta$ to the transport cross section $\sigma_{\rm tr}$ of a thermal parton in the medium. In QCD, like in QED, the interaction is, apart from Debye screening, long-rangedand thus the transport cross section is dominated by small angle scattering. Hence we find:
\begin{equation}
\sigma_{\rm tr} \approx \frac{4}{\langle\hat{s}\rangle} \int k_T^2\, dk_T^2 \frac{d\sigma}{dk_T^2}
= \frac{4\,\hat{q}}{\langle\hat{s}\rangle\rho} ,
\end{equation}
where $\langle\hat{s}\rangle \approx 18\,T^2$ denotes the average center-of-mass energy in a collision between two thermal partons. If we also make use of the relation $s/\rho \approx 4$ for the entropy per particle in a gas of massless particles, we obtain the remarkable relation \cite{Majumder:2007zh}:
\begin{equation}
\frac{\eta}{s} \approx \frac{4\,T}{5\,s\,\sigma_{\rm tr}} \approx 1.25\,\frac{T^3}{\hat{q}}
\end{equation}
This relation holds under two general conditions, which one may call the ``weak'' coupling scenario (see Fig.~\ref{fig4}):
\begin{enumerate}
\item The medium is described by nearly massless quasi-particles with the same properties (quantum numbers) as high energy partons.
\item The thermal interactions among medium partons are dominated by small-angle scattering.
\end{enumerate}
These conditions are fulfilled in a perturbative quark-gluon plasma, which can be described by the so-called hard-thermal loop effective theory; they also hold when the shear viscosity is dominated by anomalous processes in a turbulent quark-gluon plasma. They do not hold, and the relation is violated, at strong coupling, e.g.\ in the strongly coupled $N=4$ supersymmetric Yang-Mills theory, where
\begin{equation}
\frac{\eta}{s} \to \frac{1}{4\pi} \gg 1.25\,\frac{T^3}{\hat{q}} \approx \frac{1}{6\sqrt{g^2N_c}} .
\end{equation}

\section{Comparison with RHIC Data}

What do the RHIC experiments tell us about the validity of this relation for the quark-gluon plasma formed in nuclear collisions at RHIC? For this comparison we have to extract the value of the energy loss parameter $\hat{q}$ from the jet quenching data, i.e.\ from the measured suppression of the yield of energetic hadrons. A comprehensive analysis of the PHENIX data on $\pi^0$ suppression in Au+Au collisions has recently been performed by Bass {\em et al.} \cite{Bass:2008rv} in three different schemes of radiative jet quenching (ASW \cite{Salgado:2002cd}, higher twist (HT) \cite{Guo:2000nz}, and AMY \cite{Arnold:2002zm}). All calculations used the same evolution model for the medium, {\em viz.} three-dimensional ideal hydrodynamics. The Table shows the results for all three schemes, always normalized to $T=400$ MeV and for a gluon-induced jet.

\begin{table}[bht]
\begin{center}
\begin{minipage}{16.5 cm}
\label{tab:qhat}
\caption{Values of the energy loss parameter $\hat{q}$ in GeV$^2$/fm deduced \cite{Bass:2008rv} from the $\pi^0$ yield in $200\,A$ GeV Au+Au collisions at RHIC, normalized to gluons and $T=400$ MeV. The scaling of $\hat{q}$ with medium density is either assumed to be proportional to $T^3$ or to $\varepsilon^{4/3}$, where $\varepsilon$ denotes the energy density of the medium.}
\end{minipage}\\[5mm]
\begin{tabular}{|c|c|c|c|}
\hline
&&&\\[-3mm]
scaling & ASW & HT & AMY \\
&&&\\[-3mm]
\hline
&&&\\[-3mm]
$T^3$ & 10 & 2.3 & 4.1 \\
$\varepsilon^{4/3}$ & 18.5 & 4.5 & --- \\[-3mm]
&&&\\
\hline
\end{tabular}
\end{center}
\end{table}

As the Table shows, there is no consensus yet between the different schemes. For our purposes here, let us adopt the AMY scheme, because a more complete analysis has been performed in this scheme, which also includes collisional energy loss in the same framework \cite{Qin:2007rn}. The additional energy loss mechanism leads to a somewhat lower value of $\hat{q} \approx 2.75$ GeV$^2$/fm, corresponding to
\begin{equation}
1.25\, (T^3/\hat{q})_{\rm RHIC} \approx 0.145 ,
\end{equation}
which nicely falls into the range $0\leq \eta/s \leq 0.3$ preferred by the RHIC experiments. Even more impressive is the coincidence with the value $\eta/s = 0.134 \pm 33$ obtained by Meyer \cite{Meyer:2007ic} for $T = 1.65\,T_c \approx 300$ MeV in quenched lattice QCD. This suggests that the matter produced at RHIC is a partonic quasi-particle plasma. [If, on the other hand, the value of $\hat{q}$ shown in the Table for the ASW approach is correct, this would imply a value $1.25\,T^3/\hat{q} \approx 0.03-0.04$, which lies far below the KSS bound for $\eta/s\approx 0.08$ and thus would prove that the matter is a strongly coupled plasma in the true sense of the word.]

\section{Summary}

The shear viscosity $\eta$ and the radiative energy loss parameter $\hat{q}$, both probe the momentum transport properties of the medium. A small value of the (kinematic) shear viscosity $\eta/s$ corresponds to a large energy loss parameter. In other words, an almost perfect QCD fluid is extremely opaque to particles carrying color charge. A simple inverse relation holds in all thermal gauge theories, including QCD, at weak coupling. At strong coupling, the ratio $\eta/s$ is limited by the KSS bound, but $\hat{q}$ can become arbitrarily large. 

The existing approaches to jet quenching do not agree in their conclusions about the physical structure of the quark-gluon plasma produced in nuclear collisions at RHIC, some hinting at a ``weakly'' coupled plasma with quasi-particle excitations, and some suggesting a strongly coupled medium. Reliable determinations of $\eta/s$ and $\hat{q}$ from RHIC (and soon LHC) data thus have great importance.

\section*{Acknowledgments}

I thank the organizers for the invitation to the International School of Nuclear Physics in Erice, Sicily and support. This work was supported in part by a grant from the U.S. Department of Energy (DE-FG02-05ER41367).


\end{document}